\documentclass[12pt]{article}%
\usepackage{amsmath}
\usepackage{graphicx}%
\usepackage{amsfonts}%
\usepackage{amssymb}
\begin{document}

\author{C. S. Unnikrishnan$^{1}$ and G. T. Gillies$^{2}$\\$^{1}$\textit{Gravitation Group, Tata Institute of Fundamental Research, }\\\textit{\ Homi Bhabha Road, Mumbai - 400 005, India}\\$^{2}$\textit{School of Engineering and Applied Science,}\\\textit{University of Virginia}, \textit{Charlottesville, VA 22904-4746, USA}\\E-mail address: $^{1}$unni@tifr.res.in, $^{2}$gtg@virginia.edu}
\title{Rigorous Comparison of Gravimetry Employing Atom Interferometers and the
Measurement of Gravitational Time Dilation }
\date{\textit{ }}
\maketitle

\begin{abstract}
We present a gravitationally rigorous and clear answer, in the negative, to
the question whether gravimetry with atom interferometers is equivalent to the
the measurement of the relative gravitational time dilation of two clocks
separated in space. Though matter and light waves, quantum states and
oscillator clocks are quantum synonymous through the Planck-Einstein-de
Broglie relations and the equivalence principle, there are crucial differences
in the context of tests of gravitation theories.


\pagebreak 

\end{abstract}

The relative time dilation of two clocks at different locations in a
gravitational field was predicted and calculated by Einstein in 1911 using the
principle of equivalence and the Doppler shift in motion, well before the full
theory of general theory of relativity was formulated \cite{einstein11}.
Einstein's insightful interpretation of the shift in the frequency of the
radiation as a consequence of the clocks running at different rates is the
connection between the phenomenon of gravitational redshift and the
gravitation time dilation. Since frequency is a number referred to a clock,
only a change in the rate of the clock can change the frequency. Measurement
of gravitational redshift compares the frequency or energy of the radiation
between two points directly as in the Pound-Rebka experiment, whereas
gravitational time dilation is best measured by comparing the accumulated time
difference between two physical clocks in a gravitational field with radiation
or an electromagnetic contact serving only as a means of communication, needed
only at the beginning and the end of the comparison, in those cases when the
clocks cannot be transported for comparison.

An important unifying idea is that \emph{once the frequency of an oscillator
is specified, its progressive phase is equivalent to time}; there is no
difference between physical time and physical phase if an oscillator is used
as the basis of time measurement. This implies the universality of the
gravitational effect on all time dependent phenomena \cite{ug-gress11}.
Gravity couples to the total energy of a physical system, and energy is
proportional to a `frequency' in quantum physics and optics and hence the same
physical system in two different gravitational potentials have different time
evolutions and accumulated phase. Since the phase of an oscillator of every
kind is equivalent to time, states, photon, and matter waves can all be
interpreted as `clocks' in quantum mechanics, apart from real physical clocks
themselves \cite{ug-gress11}. A stationary state of definite energy in quantum
mechanics is an `oscillator', with free time evolution factor $\exp
(-iEt/\hbar),$ with a `frequency' given by the relation $\nu=E/h.$ Light obeys
the relation $E=h\nu$ in spite of being treated as discrete photons. And a
clock, like an atomic clock, is based on transitions that obey $\Delta
E=h\nu.$ However, there is a price to pay for this universality -- the
associated wave is an abstract and unobservable entity, manifesting only
through its relation to relevant probabilities, except perhaps in the case of
low frequency radiation where the interaction of the `wave' on charged
particles can be seen directly. We stress this point because it is important
in the rigorous and correct interpretation of what the measurement of time in
a gravitational field means. It is only in a space-time interpretation of
quantum physics, as in the de Broglie-Bohm theory for example, spatial
ontological status can be ascribed to the quantum wave.

The analysis of quantum dynamics in weak (laboratory) gravitational fields
reduces to the relations $E_{g}=-E\phi_{g}/c^{2}$ and the phase $d\varphi
=E_{g}(x,t)dt/\hbar$ accumulated over time duration $dt,$ where $E$ is total
energy of the physical system and the gravitational potential $\phi_{g}$ is
simply related to the metric component $g_{00}.$ The gravitational part of the
quantum evolution is determined by the integrated phase over the path given by
$\Delta_{g}(x)=\int E_{g}(x)dt/\hbar.$

The gravitational energy of a massive particle is $E_{g}=-m_{g}\phi_{g}$ and
therefore, the quantum phase depends explicitly on the gravitational mass.
However, as stressed in references \cite{harmony,exotica}, there is full
compatibility with the classical equivalence principle. Since the quantum
phase is proportional to the product of this energy and the time spent in the
potential, $t\simeq l/v,$ where $\ l$ is the spatial scale and $v$ the
velocity of the particle, the accumulated phase in each path is
\begin{equation}
\Delta_{g}\simeq E_{g}t/\hbar=-m_{g}\phi_{g}l/v\hbar\label{phase1}%
\end{equation}
We can rewrite this expression, using the relation between the inertial mass
and the de Broglie wavelength in quantum theory, $\lambda_{dB}=2\pi\hbar
/m_{i}v,$ as
\begin{equation}
\Delta_{g}=-m_{g}\phi_{g}lm_{i}\lambda_{dB}/\hbar^{2} \label{phase2}%
\end{equation}
or as
\begin{equation}
\Delta_{g}=-\left(  \frac{m_{g}}{m_{i}}\right)  \left(  \frac{\phi_{g}}{v^{2}%
}\right)  \left(  \frac{l}{\lambda_{dB}}\right)  =-\left(  \frac{E_{g}%
}{2E_{kin}}\right)  \left(  \frac{l}{\lambda_{dB}}\right)  \label{phase3}%
\end{equation}
The expression is particularly interesting due to the scaling expressed in
terms of the kinetic energy and the wavelength, and more importantly due to
the appearance of the ratio of the gravitational and the inertial mass.

For atomic clocks that work with a transition frequency $\nu$ between two
stationary states, $E=h\nu,$ the accumulated gravitational time dilation is
$\delta T=\Delta_{g}/2\pi\nu$ and the relative time dilation for \emph{two
clocks} at points $x_{1}$ and $x_{2}$ is given by
\begin{equation}
\delta T_{r}=\frac{\Delta_{g}(x_{1})-\Delta_{g}(x_{2})}{2\pi\nu}=T\left[
\phi_{g}(x_{1})-\phi_{g}(x_{2})\right]  /c^{2}=Tgl/c^{2} \label{t-dil}%
\end{equation}
Here, $l=x_{1}-x_{2}$ and $T$ is the total duration of comparison. We have
assumed that the potential over each of the relevant path is constant, for
simplicity. This is the standard general relativistic expression.

Several points about equation \ref{t-dil} is worth mentioning. The accumulated
time dilation is independent of the properties of the clock, especially its
mass etc. This is a consequence of implicitly assuming the equivalence
principle exactly, as we will show. It is also independent of the frequency of
the oscillator, for the same reason. The time dilation is the difference
between the phases accumulated over each path (or position in the
gravitational field), usually by two different clocks, each of which can serve
as an independent physical clock. If phase difference is the measured
quantity, the smallest time dilation that can be resolved, or the precision of
the measurement, is inversely proportional to the frequency of the
`oscillator', a point that is important for the discussion here.

We now show that the equation \ref{t-dil} requires a subtle modification when
the equivalence principle is not assumed in its exact form. This is relevant
for experiments that test small violations of the equivalence principle. The
gravitational coupling is the gravitational mass whereas the Shrodinger
evolution equation and relations of the form $E=mc^{2}=h\nu$ refers to the
inertial mass. Keeping this distinction in mind and referring to massive
particles, the gravitationally modified phase is
\begin{equation}
\Delta_{g}(x)=\int E_{g}(x)dt/\hbar=-\int m_{g}\phi_{g}dt/\hbar
\end{equation}
Then the expression for time dilation should be written as
\begin{equation}
\delta T_{r}=\frac{\Delta_{g}(x_{1})-\Delta_{g}(x_{2})}{2\pi\nu}=\left(
\frac{m_{g}}{m_{i}}\right)  Tgl/c^{2} \label{t-dil-mod}%
\end{equation}

The presence of the term $\Delta_{g}(x_{1})-\Delta_{g}(x_{2})$ referring to
the accumulated phase difference between two oscillator clocks is tempting
enough to interpret the differential phase accumulated in situations of
two-path interferometry as a differential clock delay. For example, the case
of a single photon in a double slit experiment in which the slits are
positioned vertically in a gravitational field involves gravitationally
induced phase difference between the two possible paths to the detector point
on screen,
\begin{equation}
\Delta_{g}=-\frac{1}{\hbar}\int dt\left(  h\nu/c^{2}\right)  \left(  \phi
_{g}(x_{1})-\phi_{g}(x_{2})\right)
\end{equation}
This is negligibly small in laboratory situations. In principle, it can be
made measurable \ as a shift of the fringe position proportional to the
gravitational field. In contrast, if massive atoms or neutrons are used in
double-slit configuration in a gravitational field, the `mass term' is
$10^{9}$ to $10^{11}$ times larger, being the ratio $mc^{2}/h\nu$. This is the
enormous advantage of matter wave interferometry in metrology, even though
some of this advantage gets lost in the fact that the extent of the path is
usually much smaller than what is possible in the case of optical
interferometry. An important question, raised by a recent debate
\cite{chu,wolf}, is whether the measurement of the gravitationally induced
phase in such an experiment can be interpreted as the measurement of
gravitational time dilation between two clocks at different points in a
gravitational field. M\"{u}ller \textit{et al} claimed that the quantum phase
difference accrued between the two quantum states of the same particle in
situations of atom interferometry, as in gravimetry with an atom
interferometer, is equivalent to the measurement of the gravitational time
dilation between two clocks \cite{chu}. In their view, the quantum phase over
each path over a duration $T$ can be re-written in terms of the Compton
frequency $\omega_{c}=mc^{2}/\hbar$ as
\begin{equation}
\Delta_{g}(x)=-m\phi_{g}(x)T/\hbar=-\frac{mc^{2}}{\hbar c^{2}}\phi
_{g}(x)T=-\omega_{c}\phi_{g}(x)T/c^{2}%
\end{equation}
The smallest time dilation factor than can be measured is then $\delta
T=\Delta_{g}/\omega_{c}$. Therefore, if it is legitimate to imagine the moving
atom as a real `Compton wave clock' in space and time, the expression for the
gravitational phase shift written in terms of the Compton frequency suggests
exceptional and unprecedented sensitivity for the measurement of gravitational
time dilation, since the Compton frequency is more than $10^{10}$ times larger
than even optical frequencies of modern atomic clocks (this gain is the same
as the ratio of the rest mass to photon energy, noted earlier). Based on this
assumption and the results of gravimetry employing atom interferometry
performed about a decade ago \cite{kasevich,chu-gravimeter}, M\"{u}ller
\textit{et al} claimed \cite{chu} enormous improvement of the measurement of
gravitational time dilation, by a factor of about 10$^{6}.$

The comment of disagreement by Wolf \textit{et al} focussed on dismissing the
notion of treating the moving atom as a Compton wave in space \cite{wolf}.
They pointed out that the expression for the measured differential phase shift
does not contain the Compton frequency, being independent of the mass of the
atom. The debate saw each side holding on to their views, supported on both
sides by similar arguments by other authors \cite{chu2,muller,sam}. We aim and
achieve a gravitationally rigorous and physically justified resolution of the
issue. We do this by proving the following main results: a) the correct
expression for phase difference in an atom interferometer involves the ratio
of gravitational and inertial masses, and also the de Broglie wavelength
$\lambda_{dB}$, and not the Compton wavelength, b) The phase shift is
essentially the ratio of free fall distance in the gravitational field $g$ and
the de Broglie wavelength, and c)\emph{ }the mass term does not drop out,
since the de Broglie wavelength $\lambda_{dB}\sim1/m$ explicitly appears.
However this can be replaced with the differential momentum $\kappa$ of a
two-photon Raman transition used in the experiment, since $\lambda_{dB}%
=2\pi/\kappa,$ to give the impression that it is independent of the mass. We
also point out two other relevant points in support; 1) a quantum system in a
superposition of two eigen states that develop a spatial separation in state
space is not equivalent to physical clocks that are spatially separated, each
of which can be accessed for comparison with a primary clock, \ 2) the fact
that it is the de Broglie wavelength that is associated with the atoms in
propagation can be directly checked in a conventional double slit experiment,
revealing the spatial fringes scaled to the de Broglie wavelength. The
associated frequency obeys the relation $E=h\nu$ with $E=p^{2}/2m$ and not
with $E=mc^{2}.$

Quantum dynamics of a system that can be in a superposition of two states can
be used for atom interferometry by entangling the states with positional
degree of freedom. In the case of atom interferometry with a two-state atom,
creating superposition of the two states involves laser excitation that
attaches different momenta to the two states, which results in evolution that
associates two different spatial positions with the two states. For example,
coherent resonant excitation by a laser with sufficient pulse duration
($\pi/2$ in terms of the inverse Rabi frequency) that creates an equal
superposition $\left(  \left|  g\right\rangle +\left|  e\right\rangle
/\sqrt{2}\right)  $ of ground and excited states $\left|  g\right\rangle $ and
$\left|  e\right\rangle $ starting with the ground state, generates the
detailed entangled state $\left|  g,0\right\rangle +\left|  e,\hbar
k\right\rangle /\sqrt{2}).$ The difference in momentum $\hbar k$ develops into
a separation of the quantum states in spatial coordinates, visible on
measurement as two distributions of atoms centred on these spatial positions,
separated by $\Delta x=\hbar kt/m$. The differential quantum phase of such
entangled states in a gravitational field is the basis of gravimeters and
inertial sensors employing atom interferometry \cite{review-atom-inter}.

Several points arising from the peculiarities of the representational space of
quantum theory are important enough to mention. First of all, there is no
consistent interpretation of each atom in the superposed state as two
spatially separated physical entities moving over different trajectories in
the gravitational field. Standard quantum theory prohibits such a view since
the quantum state has no consistent spatial ontology. While the differential
phase at an end point can be measured, a direct measurement of the spatial
position of each atom will result in one or the other spatial position. For
the slowly moving atoms, it is possible to associate a wave that obeys the
wave equation for evolution, but that is uniquely the de Broglie wave and not
the Compton wave. This fact can be directly checked by subjecting the same
atoms to a conventional double slit interference experiment and observing the
`fringes' which show the pattern corresponding the ratio of the de Broglie
wavelength $\lambda_{dB}=2\pi\hbar/m_{i}v$ to the slit separation. Hence,
there is no empirical support to associate a Compton wave to the slow atoms in
atom interferometry, \emph{except as a conversion of the unit of mass}.%

\begin{figure}
[ptb]
\begin{center}
\includegraphics[
height=1.7417in,
width=4.0006in
]%
{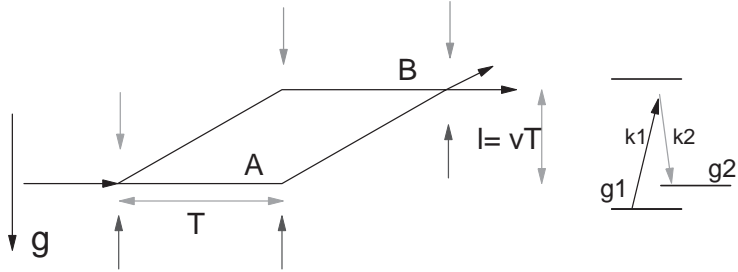}%
\caption{The space-time diagram of an atom interferometer. The laser pulses
from a pair of Raman lasers, of duration Pi/2, Pi and Pi/2 in terms of Rabi
frequency, create the superposition of hyperfine states that separate and
recombine in space due the differential momentum imparted. The differential
momentum of the states $p=\hbar(k_{1}-k_{2}).$}%
\end{center}
\end{figure}

The geometry relevant for atom interferometry involving ground states $\left|
g1\right\rangle $ and $\left|  g2\right\rangle $ with energy difference
$\Delta E/h=\nu_{0}$ and an upper state $\left|  e\right\rangle $ is indicated
in figure 1 \cite{kasevich}. A two-laser system is used for Raman pulses, each
detuned sufficiently from the excited state and adjusted to have a difference
in their frequency equal to $\nu_{1}-\nu_{2}=\nu_{0}.$ The laser pulses of
duration $\pi/2,$ relative to the inverse of the Rabi frequency $\Omega,$
create the superposition of the ground states (hyperfine states). However,
$\left|  g2\right\rangle $ is associated with an excess momentum $\hbar
k_{1}-\hbar k_{2}=\hbar\kappa$ and the entangled state `separate' by
$l=\hbar\kappa T/m$ over duration $T.$ A $\pi$ laser pulse inverts the states
and the momenta and the paths recombine after duration $T$ at which point a
$\pi/2$ pulse creates similar states at the interferometer output. The
gravitational phase difference is simply the difference in phase accumulated
over the path sections A and B with spatial separation $l$ and temporal extent
$T$, with gravitational potentials $\phi_{g}(A)$ and $\phi_{g}(B).$%

\begin{align}
\delta\Delta_{g}  &  =-m_{g}\left(  \phi_{g}(x_{1})-\phi_{g}(x_{2})\right)
T/\hbar=-m_{g}glT/\hbar\nonumber\\
&  =-m_{g}g(\hbar\kappa T/m_{i})T/\hbar=-\left(  \frac{m_{g}}{m_{i}}\right)
g\kappa T^{2} \label{ph-kappa}%
\end{align}
This is the expression for gravitational differential phase shift in
gravimetry. All other phases, dynamical and laser induced, in the problem are
equal for the two paths and drop out.

If the equivalence principle is assumed, then the ratio $m_{g}/m_{i}$ drops
out and we get the familiar expression, $\delta\Delta_{g}=-\kappa gT^{2}.$
However, the general impression that the mass drops out from this expression
is not correct since the mass term is just hidden in its relation to $\kappa.$
Since $\kappa=lm/\hbar T$ and $l=vT$ where $v$ is the relative velocity
between the two wave packets, $\kappa=mv/\hbar=2\pi/\lambda_{dB}.$ The
two-photon momentum difference replaces its equivalent relative momentum of
the wave-packets, which is mass dependent. Here $\lambda_{dB}$ is a reduced de
Broglie wavelength, $1/\lambda_{dB}=1/\lambda_{dB1}-1/\lambda_{dB2}.$ The
differential phase shift can be written as
\begin{equation}
\delta\Delta_{g}=-\kappa gT^{2}=-\frac{2\pi gT^{2}}{\lambda_{dB}}
\label{phase-deB}%
\end{equation}
\emph{The phase shift is essentially the ratio of free fall distance in the
gravitational field }$g$\emph{ and the de Broglie wavelength,} as dictated by
the equivalence principle. In a picture projected to real space, the fringes
from the interference of the de Broglie waves will `fall' through a distance
$2gT^{2}$ over time $2T$ and the phase shift is the ratio of this fringe shift
and the `centre of mass de Broglie wavelength' $h/(mv)/2=2\lambda_{dB}.$ It is
clear that the gravitational phase is scaled to the de Broglie waves, as
expected for slow non-relativistic atoms, and not to the relativistic and
notional Compton wave.

Every expression in quantum theory that involves the nonrelativistic de
Broglie wavelength $\lambda=h/mv$ can be re-written in terms of the Compton
wavelength with a change of unit from mass to frequency, as
\begin{equation}
\lambda=h/mv=\frac{hc^{2}}{mc^{2}v}=\frac{c^{2}}{\omega_{c}v}=\frac{c^{2}%
T}{\omega_{c}l}%
\end{equation}
This is the Compton wavelength multiplied by the large factor $cT/l,$ the
ratio of the light travel distance to the spatial separation between the
atomic states, reiterating the fact that it is not the Compton wave that is
relevant in the experiment and the measurement. Then the relative phase shift,
assuming the equality of the inertial and gravitational masses, is
\begin{equation}
\delta\Delta_{g}=-\omega_{c}glT/c^{2}%
\end{equation}
The appearance of \ the `relativistic' Compton frequency is therefore
fictitious. \ The explicit dependence of phase is on the de Broglie wavelength
and not on the Compton wavelength, as revealed in equation \ref{phase-deB}.
This can of course be empirically proved simply by forming a spatial
interference pattern with the same slow atoms through a double slit, revealing
the underlying relevant wavelength in the spatial interference pattern.

We stress the important point that a physical clock should admit standard
clock operations relative to a primary standards and for this it is necessary
that the oscillator phase is directly accessible for comparison. For the atom
interferometer gravimeter, this phase is manifested in the population of
either of the hyperfine states after recombination, determined by the
\emph{phase difference} imprinted gravitationally due to the difference in the
interaction energy, $-m(\phi_{g}(x_{1})-\phi_{g}(x_{2})).$ Hence we cannot
treat each of the individual wavepackets as individual clocks, just as the
two-state quantum superposition separated in a Stern-Gerlach magnet is not two
individual physical systems. They do not even exist in space as physical
reality, except in certain non-standard interpretations of quantum mechanics.

What constitutes a genuine clock and what is its ultimate precision in a
measurement of gravitational time dilation is a hard problem in the context of
the apparent universal behaviour of waves, quantum states and physical clocks
in a gravitational field. Our analysis helps to provide a satisfactory answer
that is rigorous in its treatment of the equivalence principle and the effect
of weak gravitational fields and goes a long way in clarifying several issues.
We have established that the gravitational phase shift in atom interferometry
involves the non-relativistic momentum and the de Broglie wave rather than the
Compton wave. The relation to the equivalence principle is brought out clearly
by demonstrating how the ratio of the gravitational mass to inertial mass
remains in the equations for the differential quantum phase. Moreover, we have
argued that the observable oscillator is not the Compton wave because the
relevant spatial fringe pattern is determined by the de Broglie wavelength.
The Compton wave has no physical manifestation in gravimetry with massive
particles and the Compton frequency in the expression for phase shift is no
more than a conversion of the unit of mass.

\end{document}